\RequirePackage{fix-cm}

\documentclass[twocolumn,epjc3]{svjour3}          

\RequirePackage[T1]{fontenc}

\smartqed  

\RequirePackage{graphicx}
\RequirePackage{mathptmx}      
\RequirePackage{flushend}
\RequirePackage[numbers,sort&compress]{natbib}
\RequirePackage[colorlinks,citecolor=blue,urlcolor=blue,linkcolor=blue]{hyperref}

\usepackage{enumerate}
\usepackage{amssymb}
\usepackage{amsmath}
\usepackage{amsfonts}
\usepackage{bm}
\usepackage{bbm}

\journalname{Eur. Phys. J. C}

\begin{document}

\title{Localization of abelian gauge fields on thick branes}

\author{Carlos A. Vaquera-Araujo\thanksref{addr1,e1}
        \and
        Olindo Corradini\thanksref{addr2,addr3,e2} 
}

\thankstext{e1}{e-mail: carolusvaquera@gmail.com}
\thankstext{e2}{e-mail: olindo.corradini@unach.mx}

\institute{Facultad de Ciencias, CUICBAS, Universidad de Colima, Bernal D\'iaz del Castillo 340, Colima 28045, M\'exico \label{addr1}
          \and
          Facultad de Ciencias en F\'isica y Matem\'aticas, Universidad Aut\'onoma de Chiapas, Ciudad Universitaria, Tuxtla Guti\'errez 29050, M\'exico\label{addr2}
          \and
          Dipartimento di Scienze Fisiche, Informatiche e Matematiche, Universit\`a di Modena e Reggio Emilia, Via Campi 213/A, Modena I-41125, Italy\label{addr3}
}

\date{}

\maketitle

\begin{abstract}
In this work, we explore a mechanism for abelian gauge field localization on thick branes based on a five-dimensional Stueckelberg-like action. A normalizable zero mode is found through the identification of a suitable coupling function between the brane and the gauge field. The same mechanism is studied for the localization of the abelian Kalb--Ramond field.
\end{abstract}

\section{Introduction}
In the brane-world scenario, our universe is depicted as a four-dimensional (4D) sub-manifold (3-brane) embedded in a 
higher dimensional space-time. \,
Within this framework, gravity is able to propagate in all dimensions, but matter fields are restricted to live on the 3-brane. 
Among the most attractive schemes constructed under this hypothesis stand the proposals of Randall and Sundrum \cite{Randall:1999ee,Randall:1999vf} (resp. RS1, RS2), which involve only one extra dimension and a non-trivial warp factor due to the underlying anti-de Sitter (AdS) geometry. 

Branes in RS models and their generalizations are very idealized, they are introduced as infinitely thin (singular) hyper-surfaces. Besides, such thin branes are static, with no dynamical mechanism responsible for their formation. In order to avoid the use of singular branes, thick branes modeled as domain wall configurations can be implemented in extra dimensional theories (for a detailed review on thick brane solutions, see \cite{Dzhunushaliev:2009va}). The key feature of this approach is that thick branes are dynamically generated by one or several background scalar fields coupled with gravity. 

In the pursuit of a more realistic picture of the brane-world, it is also important to provide a natural localization mechanism of bulk fields on domain walls. As shown in \cite{Gremm:1999pj,Kakushadze:2000zp}, the graviton can be successfully localized on a thick brane built with a single background scalar in an asymptotic five-dimensional (5D) AdS space-time. Massless scalar fields can also be localized in this minimal setup, but unfortunately vector gauge fields seem to require a richer brane structure. The essence of this ``no-go theorem'' is already captured  in the thin brane limit (RS2 model)~\cite{Bajc:1999mh}.

A significant amount of work has been devoted to the problem of vector gauge field localization in the context of 5D AdS space-time, in both  singular and thick domain wall branes. The models available in literature show a wide variety of ideas, which include: mass terms for the vector boson \cite{Ghoroku:2001zu,Kogan:2001wp,Batell:2005wa},  coupling between the dilaton and the kinetic term of gauge fields \cite{Kehagias:2000au,Cruz:2010zz}, insertion of kinetic terms induced by localized fermions \cite{Guerrero:2009ac}, and a smearing out dielectric function inspired by the Friedberg--Lee model for hadrons \cite{Chumbes:2011zt}. Recently, new mechanisms of gauge field localization have also been  studied in diverse brane-world models, including  Weyl thick branes \cite{Liu:2008wd}, two-field thick branes \cite{Fu:2011pu}, AdS 3-branes \cite{Liu:2011zy}, and tachyonic domain walls \cite{Herrera-Aguilar:2014oua}.

The inherent difficulty of trapping gauge fields in a domain wall is generically present for all antisymmetric form fields, such as for instance the Kalb--Ramond (KR) tensor, which arises as a massless bosonic mode in closed string theories and can act as a source of torsion in a Riemannian manifold \cite{Mukhopadhyaya:2002jn}. Any chance of localizing the KR field can be useful to determine the observable effects of torsion in the 3-brane. The corresponding zero mode of the KR tensor in five dimensions is known to be non-localizable on the simplest thick brane, generated by a single real scalar field \cite{Tahim:2008ka}. As with its gauge vector counterpart, the problem concerning the localization of the KR field has been considered by several authors~\cite{Chumbes:2011zt,Tahim:2008ka,Cruz:2013zka,Du:2013bx,Liu:2011ysa,Christiansen:2010aj}.

In this work, a new mechanism for gauge field localization on thick branes is explored. The model can be seen as a domain wall  version of the mechanisms presented in \cite{Ghoroku:2001zu,Batell:2005wa}, within a regularized RS2 scenario.  The basic setup of the model is a thick brane embedded in an asymptotically 5D AdS bulk space-time, described by a single real background scalar field. In this framework, localization is achieved by the introduction of Stueckelberg-compensating fields in the 5D action of a gauge field, where the quadratic coefficient of the gauge fields is modeled as a Yukawa-like brane--gauge coupling. The second part of the paper is devoted to showing that the same mechanism can be readily generalized to localize the zero mode of the Stueckelberg-like KR field, through a suitable modification of the Yukawa coupling. It is important to notice that the use of Stueckelberg-compensating fields restricts the application of this mechanism to abelian gauge fields.

The structure of the paper is the following: In Sect. \ref{Setup} a brief review of the brane-world
generated by a single real scalar is presented. In Sect. \ref{VGFL} it is shown that starting from the Stueckelberg-compensated action of a vector gauge field, a normalizable zero mode can be localized on the brane whenever a suitable brane--gauge coupling function is fixed. A similar procedure to localize the zero mode of the KR field is introduced in Sect. \ref{KRL}. Finally, the conclusions of the work are contained in Sect. \ref{Conc}.

\section{Thick brane generated by a single scalar}\label{Setup}
Consider a model of 5D gravity coupled to a single scalar field $\phi$:
\begin{equation}
  S = M_{*}^3 \int d^5x \sqrt{-g} \left[R-\frac12
 g^{MN} \partial_M \phi\, \partial_N \phi-V(\phi)\right]\, ,
\label{eq:model}
\end{equation}
with $M_{*}$ being the fundamental Planck scale in five dimensions and $V(\phi)$ the scalar potential. The scalar field has been rescaled in units of $M_{*}$, such that $\phi$ is dimensionless and $V(\phi)$ has mass dimension 2. We are interested in background solutions  where the metric displays 4D Poincar\'e symmetry
\begin{equation}
  ds^2 = g_{MN} dx^M dx^N= e^{2A(y)} \eta_{\mu\nu} dx^\mu dx^\nu +dy^2\, . 
\end{equation}
Here the 5D coordinates $x^M$ ($M=0,\dots,4$) are separated into 4D Minkowski space-time coordinates $x^\mu$ ($\mu=0,\dots,3$), with $\eta_{\mu\nu}=\text{diag}(-1,+1,+1,+1)$, and the infinitely extended extra dimension $y$. Assuming that the scalar field $\phi$ depends exclusively on $y$, the equations of motion of Eq.~(\ref{eq:model}) become
\begin{eqnarray}
&&\label{eq:p}  \phi'' + 4 A'\phi' =\partial_\phi V ,\\
&&\label{eq:55} 12 {A'}^2 -\frac12 {\phi'}^2 +V =0 ,\\
&&\label{eq:mn-55} 3A''+\frac12 {\phi'}^2 =0 , 
\end{eqnarray}
where the prime denotes derivative with respect to $y$.

According to the super-potential method~\cite{Gremm:1999pj}, a solution for Eqs.~(\ref{eq:p})--(\ref{eq:mn-55}) can be found through the first-order differential equations
\begin{equation}\label{eq:solpA}
 \phi' = \partial_\phi W ,\qquad  A' = -\frac16  W ,
\end{equation}
if the scalar potential is written in terms of some super-potential function $W(\phi)$ as
\begin{equation}\label{eq:solV}  
V = \frac12(\partial_\phi W)^2 -\frac13  W^2\, . 
\end{equation}
An example of such super-potential is given by the sinusoidal function \cite{Gremm:1999pj}
\begin{align}
W(\phi) = 6 a b \sin \left(\frac{\phi }{\sqrt{6 b} }\right)\, ,
\end{align} 
where $b>0$. This functional form of the super-potential determines the following background solution for the scalar field: 
\begin{align}
\phi(y) &= 2 \sqrt{6 b} \arctan\left[\tanh\frac{a (y-y_0)}{2}\right]\, ,\label{phis}
\end{align}
which yields a domain wall with positive tension
\begin{align}
T = M_{*}^3\int_{-\infty}^{+\infty} dy~ (\phi')^2 = 12 M_{*}^3 b |a| ,
\end{align} 
and a smooth warp factor with
\begin{align}
A(y) &= A_0-b \log [\cosh a(y-y_0)]\, .\label{As}
\end{align}
In the above equations, $y_0$ denotes the center of the domain wall and $A_0$  determines the value of the warp factor at this point. Without loss of generality, we can set $y_0=0$, $A_0=0$.

The background described above is asymptotically Anti-de Sitter $A(|y|\to \infty) \sim
- k|y|$ with AdS scale $k= b |a|$. Therefore, it constitutes a regularized version of the RS2 model. 
It is often convenient to work in a conformally flat frame, defined by the transformation $dy=e^{A}dz$, where the metric takes the form
\begin{equation}
  ds^2 =e^{2A(z)}( \eta_{\mu\nu} dx^\mu dx^\nu +dz^2)\, .\label{ccoo}
\end{equation}
Notice that this transformation cannot be written in closed analytical form for arbitrary values of the parameters associated with brane thickness ($a$) and curvature ($b |a|$) in Eqs.~(\ref{phis}), (\ref{As}). However, particular cases are tractable. For example, when $b=1$ the warp factor reads
\begin{align}
A(z) &=-\frac12\log [1+a^2z^2]\, .\label{Asz}
\end{align}
 In this work we do not adopt a particular form of the background, but only assume the existence of a domain-wall solution generated by some super-potential $W(\phi)$ with asymptotic AdS behavior $A(|y|\to \infty) \sim- k|y|$ and well defined transformation $dy=e^{A}dz$.  We refer to Eq.~(\ref{Asz}) for a concrete realization of this setup.

\section{Abelian gauge vector field localization}\label{VGFL}

Our starting point is the  Stueckelberg-like 5D $U(1)$ gauge field action
\begin{equation}
S_A=\int d^5x \sqrt{-g}\left\{-\frac{1}{4}F^{MN}F_{MN}-\frac{1}{2}\mathcal{G}(\phi)\left[\partial_M B-A_M\right]^2\right\}\, . \label{2}
\end{equation}
Here $A_M$ is the 5D gauge vector field and $B$ is a dynamical scalar field (see~\cite{Ruegg:2003ps} for a review of the Stueckelberg field). The main advantage of implementing this action is its gauge invariance in five dimensions under the transformation
\begin{equation}\label{gt1}
A_M \rightarrow  A_M+\partial_M \Lambda,\qquad
B \rightarrow B+ \Lambda\,.
\end{equation} 
The coupling of the gauge field and the brane is described by a Yukawa-like interaction $\mathcal{G}(\phi)$. 
The aim of this section is to find if the model defined by $S_A$ can lead to the localization of a gauge field zero mode on thick branes, through the adoption of a particular functional form for $\mathcal{G}(\phi)$. 

From the 4D low-energy perspective, the model defined by Eq.~(\ref{2}) can be understood in
terms of massless and massive sectors. The localization mechanism is expected to 
give rise to a massless sector, from the zero mode of $S_A$, and massive sectors coming from
the continuum of non-zero modes.\footnote{Strictly speaking, since the $y$ direction is non-compact, there is no finite gap between massless and massive modes and the massive spectrum forms a continuum. However, the transverse invariant length is finite, which should ensure a low-energy localization, along the lines of what happens in the RS2 model.}  The counting of degrees of freedom (d.o.f.) goes as
follows: for a massless $p$-form, in $d$ dimensions, one has $\binom{d-2}{p}$ d.o.f., whereas for a massive $p$-form one has $\binom{d-1}{p}$. The present model in its 5D form contains a total of 4 on-shell d.o.f.  Thus, matching  that number with 4D fields, we naively expect to obtain a 4D effective theory with a massless sector composed of a 1-form
(2 on-shell d.o.f.) plus two real scalar fields (1 on-shell d.o.f. each), and massive sectors made of a
1-form (3 on-shell d.o.f.) and one real scalar.

\subsection{Equations of motion and gauge fixing}

Varying the action $S_A$, we obtain the following 5D equations of motion for $A_M$ and $B$: 
\begin{eqnarray}
&&\partial _{M}\Big[e^{4A}g^{ML}g^{NP}F_{LP}\Big]=-e^{4A}\mathcal{G}(\phi)g^{NP} \left(\partial_P B-A_P\right)\,,  \label{5deom1}\\
&&\partial _{M}\Big[e^{4A}\mathcal{G}(\phi)g^{ML}\left(\partial_L B-A_L\right)\Big]=0\,, \label{5deom2}
\end{eqnarray}
where the equation of motion for $B$ (cf. Eq.~\eqref{5deom2}) is consistent with the Noether identity obtained by taking the divergence of Eq.~\eqref{5deom1}.
 
For simplicity, in the present work we assume that the back-reaction of $\mathcal{G}(\phi)$ into the geometry is negligible, and thus, the minimum energy solutions for	 $\phi(y)$ and $A(y)$ are determined by Eq.~(\ref{eq:solpA}). 

The first important task in our analysis is to fix the gauge. Instead of imposing an explicit gauge fixing condition, we perform an analogous analysis to the one done in  \cite{Ghoroku:2001zu,Batell:2005wa}, where the 5D field $A_M$  is parameterized as
\begin{equation}
A_M=( A_\mu, A_4)=( \widehat{A}_\mu+\partial_\mu \varphi, A_4)\,,
\label{par}
\end{equation}
with $\widehat{A}_\mu$ and $\varphi$ as the transverse ($ \partial^\mu\widehat{A}_\mu=0$) and longitudinal components of $A_\mu$,  respectively. 
The behavior of these components under the gauge transformation Eq.~(\ref{gt1}) is
\begin{equation}\label{gt2}
\begin{array}{ccc}
\widehat{A}_\mu \rightarrow \widehat{A}_\mu,&\qquad &A_4\rightarrow A_4+\Lambda'\,,\\
\varphi\rightarrow\varphi+\Lambda, &\qquad & B \rightarrow B+ \Lambda.
\end{array}
\end{equation}
These transformations suggest that it is potentially useful to redefine the scalar degrees of freedom as
\begin{equation}\label{par2}
\lambda=A_4-\varphi',\qquad
\rho=B-\varphi\,,
\end{equation}
such that, under Eq.~(\ref{gt2}), the new fields $\lambda$ and $\rho$ remain invariant:
\begin{equation}\label{gt3}
\lambda\rightarrow \lambda,\qquad
\rho\rightarrow \rho\,.	
\end{equation}
Therefore, the parameterization defined by Eqs.~(\ref{par},\ref{par2}) is roughly equivalent to the gauge fixing condition $\partial_\mu A^{\mu}=0$, but incorporates the advantage of working directly with gauge-invariant fields.

Let us now write the 5D equations of motion in terms of the components $\widehat{A}_{\mu}$, $\lambda$ and $\rho$. In first place, taking $N=\nu$ in Eq.~(\ref{5deom1}), we obtain
\begin{equation}
\left[\square+ e^{2A}\left(\partial^2_y+2A'\partial_y-\mathcal{G}\right)\right]\widehat{A}_{\nu}=\partial_\nu\left[\partial_y(e^{2A}\lambda)-e^{2A}\mathcal{G}\rho\right]\,, \label{eom0}
\end{equation}
with $\square=\eta^{\mu\nu}\partial_{\mu}\partial_\nu$.
The left-hand side of this equation is purely transverse, while its right-hand side is purely longitudinal. Thus, each side must vanish independently for a non-trivial solution.  Taking this fact into account, in this parameterization, Eqs.~(\ref{5deom1},\ref{5deom2}) become equivalent to the following system of equations:
\begin{eqnarray}
&&\left[\square+ e^{2A}\left(\partial^2_y+2A'\partial_y-\mathcal{G}\right)\right]\widehat{A}_{\nu}=0\,, \label{eom1}\\
&&\partial_y(e^{2A}\lambda)-e^{2A}\mathcal{G}\rho=0\,, \label{eom4}\\
&&e^{2A}\square\lambda+e^{4A}\mathcal{G}\left(\rho'-\lambda\right)=0\,,  \label{eom2}\\
&&e^{2A}\mathcal{G}\square\rho+\partial_y\left[e^{4A}\mathcal{G}\left(\rho'-\lambda\right)\right]=0\,.  \label{eom3}
\end{eqnarray}
Note also that the decoupling condition Eq.~(\ref{eom4}) is already satisfied in a weaker form by a combination of~(\ref{eom2}) and~\eqref{eom3}~\footnote{For a  4D massive vector theory Eq.~\eqref{eom4} corresponds to obtaining  $(\square-m^2)\hat A_\nu=0$, starting from the Stueckelberg description.}:
\begin{equation}
\square\left[\partial_y(e^{2A}\lambda)-e^{2A}\mathcal{G}\rho\right]=0\,.\label{eom5}
\end{equation}

Further insight can be gained writing $S_A$ in gauge-invariant components. Substituting Eqs.~(\ref{par}), (\ref{par2}) into Eq.~(\ref{2}), it can be shown that the transverse vector $\widehat{A}_\mu$ decouples from the 
scalar fields (up to vanishing surface terms) and the action becomes
\begin{eqnarray}
S_A&=&S_{\widehat{A}}+S_S\,,\\
S_{\widehat{A}}&=&\int d^5 x\left\{ -\frac{1}{4}\widehat{F}^2_{\mu\nu}
-\frac{1}{2}e^{2A}(\widehat{A}'_\mu)^2-\frac{1}{2}e^{2A}\mathcal{G} 
\widehat{A}_\mu^2\right\}\,,\label{vector}\\
S_S&=&\int d^5x  \left\{ -\frac{1}{2}e^{2A}\left(\partial_\mu 
\lambda\right)^2-\frac{1}{2}e^{2A}\mathcal{G}\left(\partial_\mu \rho\right)^2\right.\nonumber\\&&\left.-\frac{1}{2}e^{4A}\mathcal{G} (\lambda-\rho')^2 \right\}\,.
\label{scalar2}
\end{eqnarray}
The equations of motion for $S_{\widehat{A}}$ are given by Eq.~(\ref{eom1}), while Eqs.~(\ref{eom2}), (\ref{eom3}) are the corresponding equations of motion for $S_{S}$.  

\subsection{Brane coupling and gauge field localization}
Once the gauge has been fixed, we can study the role played by the brane--gauge coupling $\mathcal{G}$ in the gauge vector field localization. Decomposing the gauge field as
\begin{equation}
\widehat{A}^{\mu }=\sum_{n}a_{n}^{\mu }(x)\alpha _{n}(y), \label{dec1}
\end{equation}
Equation (\ref{eom1}) reduces to
\begin{equation}
\left[\partial_y^2+ 2A'\partial_y-\mathcal{G}\right]\alpha_n(y)=-e^{-2A}m^2_n\alpha_n(y),  \label{prof1}
\end{equation}
with $\square  a_{n}^{\mu }(x)=m^2_n a_{n}^{\mu }(x)$.

In order to proceed further, it is now necessary to specify the functional form of the brane--gauge coupling $\mathcal{G}$. At this point we do not attempt to study the most general form of $\mathcal{G}$, but instead we investigate if there is a particular choice for this function endowed with physical significance that ensures the existence of a normalizable zero-energy ground state for the gauge field. In this regard, our choice for the coupling is guided in first place by the observation that the Ghoroku--Nakamura mass term \cite{Ghoroku:2001zu} successfully localizes a gauge field in a singular brane, so our Yukawa coupling must reproduce that behavior in the thin brane limit. Secondly, our coupling must be of scalar nature and it must have its origin in the underlying geometry. In our minimal setup there are two scalars readily available for the construction of $\mathcal{G}$: the scalar field $\phi$, responsible for the brane formation, and the scalar curvature $R$. We adopt  $\phi$ as the basic building block of the Yukawa-like coupling, having in mind that although a coupling dependent on $R$ can have a different interpretation \cite{Zhao:2014iqa, Alencar:2014moa}, it can be effectively modeled on the same footing by the minimum energy solution of $\phi$, as the functional
\begin{equation}\label{Rphi}
\mathcal{R}(\phi)= \frac{4}{3}(\partial_{\phi }W(\phi))^2-\frac{5}{9} W(\phi)^2
\end{equation} 
is numerically equivalent to the scalar curvature in this context:
\begin{equation}
R=-4\left[5(A')^2+2A''\right]\, .
\end{equation}

As a first approach, we adopt the following set of requirements for the construction of the Yukawa brane--gauge coupling:
\begin{enumerate}[(a)]
\item  $\mathcal{G}(\phi)$ is an even function of $y$,
\item it has mass dimension 2,
\item for simplicity, we assume that the coupling is determined by the super-potential $W(\phi)$ and its derivatives.
\end{enumerate}
From Eq.~(\ref{eq:solV}), it is evident that the potential $V(\phi)$ --which can be written as a combination of $W(\phi)^2$ and $[\partial_\phi W(\phi)]^2$-- does indeed satisfy these three requirements. Thus, taking $V(\phi)$ as a guideline, we propose the following Ansatz for the brane--gauge field coupling:
\begin{equation}
\mathcal{G}_{c_1,c_2}(\phi)=-\frac{c_1}{6}[\partial_\phi W(\phi)]^2+\frac{c_2}{36} W(\phi)^2 \,,  \label{G1}
\end{equation}
where $c_1,c_2$ are arbitrary real constants. 
Using Eqs.~(\ref{eq:mn-55}), (\ref{eq:solpA}), this functional can be written as
\begin{equation}
\mathcal{G}_{c_1,c_2}[\phi(y)]=c_1 A''(y)+c_2 [A'(y)]^2.  \label{G2}
\end{equation}
Plugging this functional into Eq.~(\ref{prof1}) and switching to the conformally flat frame defined in Eq.~(\ref{ccoo}), the equation for the mode profiles becomes
\begin{equation}
\left[\partial_z^2+\dot{A}\partial_z-c_1 \ddot{A}-\left(c_2-c_1\right) \dot{A}^2\right]\alpha_n(z)=-m^2_n\alpha_n(z)\,,  \label{prof2}
\end{equation}
where the dot denotes derivative with respect to $z$.
This equation can be cast into the form of a typical quantum mechanical problem through the rescaling
\begin{equation}
\alpha_n(z)=e^{-A(z)/2}\psi_n(z)\,,
\end{equation}
where the auxiliary wavefunction $\psi_n$ satisfies the Schr\"odinger equation
\begin{equation}
\left[-\partial_z^2+U(z)\right]\psi_n(z)=m^2_n\psi_n(z)\,,  \label{prof3}
\end{equation}
with QM potential
\begin{equation}
U(z)=\left(c_1+\frac12\right) \ddot{A}(z)+\left(c_2-c_1+\frac14\right) [\dot{A}(z)]^2\,.
\end{equation}

As a last step, in order to guarantee an effective 4D theory with a normalizable zero-energy ground state, we require the Schr\"odinger equation~(\ref{prof3}) to be rewritten as 
\begin{equation}
Q_\xi^\dagger Q_\xi\psi_n(z)=m^2_n\psi_n(z),  \label{prof4}
\end{equation}
which is of the form of a supersymmetric quantum mechanics problem, with
\begin{equation}\label{qdef}
Q_\xi=-\partial_z+\left(\xi+\frac12\right) \dot{A},\qquad Q_\xi^\dagger =\partial_z+\left(\xi+\frac12\right) \dot{A}\,,
\end{equation}
for a positive real parameter $\xi$. This restriction imposes the following tuning among the constants $c_1$, $c_2$ and $\xi$:
\begin{equation}
c_1 =\xi,\qquad c_2 =\xi^2+2\xi\,,
\end{equation}
such that the functional $\mathcal{G}[\phi(y)]$ and the QM Potential $U(z)$ become 
\begin{equation}
\mathcal{G}[\phi(y)]=\mathcal{G}_\xi[\phi(y)]\equiv\xi\left[(\xi+2)[A'(y)]^2+A''(y)\right]\,,  \label{G4}
\end{equation}
\begin{equation}
U(z)=U_\xi(z)\equiv\left(\xi+\frac12\right)^2 [\dot{A}(z)]^2+\left(\xi+\frac12\right) \ddot{A}(z)\,.\label{Uxi}
\end{equation}
Note that in the thin brane limit, our choice for the Yukawa coupling $\mathcal{G}_\xi$ coincides with the Ghoroku--Nakamura 5D mass term \cite{Ghoroku:2001zu, Batell:2005wa}.

For a background like that of Eq.~(\ref{Asz}), the masses of the modes are distributed in a continuous spectrum, as $U_\xi(z)\to 0$ when $|z|\to\infty$.
The hermiticity and positive definiteness of $Q_\xi^\dagger Q_\xi$ in Eq.~(\ref{prof4}) ensure that no normalizable negative energy modes are allowed. On the other hand, the zero-energy wavefunction annihilated by $Q_\xi$ is normalizable
\begin{equation}
\psi_0(z)=k_0e^{(\xi+\frac12)A(z)},  \label{wfzm1}
\end{equation}
and the corresponding zero mode profile $\alpha_0(y)$ turns out to be
\begin{equation}
\alpha_0(y)=k_0 e^{\xi A(y)},\label{pzmsol1}
\end{equation}
with $k_0$ as a normalization constant.

\subsection{Scalar sector}
Now we turn our attention to the scalar sector of the model. Let us first analyze in detail the localization of the zero modes. We start by decomposing the scalar fields as
\begin{equation}
\lambda=\sum_{n}\lambda_{n}(x)\beta_{n}(y),\qquad
\rho=\sum_{n}\rho_{n}(x)\gamma_{n}(y)\,,  \label{dec2}
\end{equation}
with
\begin{equation}
\square\lambda_n=m_{Sn}^2\lambda_n,\qquad \square\rho_n=m_{Sn}^2\rho_n\,.
\end{equation}
It can be shown, using Eqs.~(\ref{eom2}), (\ref{eom3}), that the scalar zero modes satisfy 
\begin{equation}
\lambda_{0}(x)\beta_{0}(y)=\rho_{0}(x)\gamma'_{0}(y)\,. \label{zmscal}
\end{equation}
From this relation, we observe that the 4D profiles $\lambda_{0}(x)$ and $\rho_{0}(x)$ are constrained to be proportional in order to have a non-trivial solution. Thus, there is only one independent scalar d.o.f. in zero mode of the scalar sector, instead of the two expected, and its localization properties are determined by the two profiles $\beta_{0}(y)$ and $\gamma_{0}(y)$.
Substituting Eq.~(\ref{zmscal}) into Eq.~(\ref{eom4}), we obtain the following equation for the zero mode profile $\gamma_{0}(y)$:
\begin{equation}
\gamma''_{0}(y)+2A'\gamma'_{0}(y)-\mathcal{G}_\xi\gamma_{0}(y)=0\,, 
\end{equation}
which coincides with Eq.~(\ref{prof1}) in the massless case.
Therefore, for $\xi>0$, there is only one normalizable solution, given by 
\begin{equation}
\gamma_0(y)=k_0 e^{\xi A}.\label{pzmsol2}
\end{equation}
Plugging this solution and Eq.~(\ref{zmscal}) back in Eq.~(\ref{scalar2}), it is clear that there is only one massless scalar in the low-energy spectrum of the theory, with effective action 
\begin{equation}
S^{0}_S= \int d^4x \left[-\frac{1}{2}\left(\partial_\mu \rho_{0}\right)^2 \right] \int_{-\infty}^{+\infty} dy\, e^{2A}\left[ (\gamma'_{0})^2+\mathcal{G}_\xi\gamma_{0}^2\right]\,.
\label{scalar3}
\end{equation}
However, a straightforward calculation shows that the integral over $y$ vanishes
\begin{equation}\label{int1}
\int_{-\infty}^{+\infty} dy\, e^{2A}\left[(\gamma_0')^2+\mathcal{G_\xi}\gamma_{0}^2 \right]= k_0\xi\left.A'e^{2(\xi+1)A}\right|_{-\infty}^{+\infty}=0\,.
\end{equation} 
Thus, despite being apparently normalizable, the scalar zero mode fails to be localized in the brane. It turns out to be a null state. This feature could be pathological and deserves further investigation, which will be presented in a future work.

Let us focus now on the general equations for the mode profiles. The decoupling condition Eq.~(\ref{eom4}) implies that $\rho_n$ can be eliminated in favor of $\lambda_n$ for all massive and massless modes as
\begin{equation}\label{subc3}
\mathcal{G}_\xi\rho_n \gamma_n=e^{-2A}\lambda_n\partial_y(e^{2A}\beta_n).
\end{equation}
Again, a non-vanishing solution requires that there is only one scalar d.o.f. contained in each mode.
Combining this relation with Eq.~(\ref{eom2}) and using  Eq.~(\ref{dec2}), the equations for the mode profiles become
\begin{align}
&\left\{\partial^2_y-\left[\left(\log\mathcal{G}_\xi\right)'+2A'\right]\partial_y -\mathcal{G}_\xi\right\}\left(e^{2A}\beta_{n}\right)\nonumber\\&\quad=-m_{Sn}^2\beta_{n}\, ,  \label{profzm3}
\end{align}
and
\begin{align}
&\left\{\partial^2_y+\left\{2A'+\frac{m_{Sn}^2\left[\left(\log\mathcal{G}_\xi\right)'+2A'\right]}{e^{2A}\mathcal{G}_\xi-m_{Sn}^2}\right\}\partial_y -\mathcal{G}_\xi\right\}\gamma_n\nonumber\\&\quad=-e^{-2A}m_{Sn}^2\gamma_n\, .  \label{profzm4}
\end{align}
The above equations can be written in Schr\"odinger form by defining the appropriate wavefunctions $\psi^\lambda_n(z)$ and $\psi^\rho_n(z)$ through
\begin{eqnarray}
\beta_{n}(z)&=&\mathcal{G}_\xi^{1/2}e^{-A/2}\psi^\lambda_n(z)\, ,\\
\gamma_{n}(z)&=&\left(\frac{\mathcal{G}_\xi-e^{-2A}m_{Sn}^2}{\mathcal{G}_\xi}\right)^{1/2}e^{-A/2}\psi^\rho_n(z)\, ,
\end{eqnarray}
such that their corresponding potentials become
\begin{align}
U^\lambda(z)=&\frac{\left(3\dot{A}+\partial_z\log\mathcal{G}_\xi \right)^2}{4}-\frac{\left(3\ddot{A}+\partial^2_z\log\mathcal{G}_\xi \right)}{2}\nonumber\\&+e^{2A}\mathcal{G}_\xi\, ,
\end{align}
\begin{align}
&U^\rho(z)=\frac{1}{4}\left[\dot{A}+\partial_z\log\left(\frac{\mathcal{G}_\xi}{\mathcal{G}_\xi-e^{-2A}m_{Sn}^2}\right) \right]^2\nonumber\\&\quad+\frac{1}{2}\left[\ddot{A}+\partial^2_z\log\left(\frac{\mathcal{G}_\xi}{\mathcal{G}_\xi-e^{-2A}m_{Sn}^2}\right) \right]+e^{2A}\mathcal{G}_\xi\, .
\end{align}

Summarizing, the model defined by Eqs.~(\ref{2}), (\ref{G4}) contains a normalizable zero-energy ground state described by a massless 1-form in four dimensions and a continuous tower of massive 1-forms and massive real scalars.   There is one additional massless scalar with apparently normalizable profile in the spectrum, but according to our analysis, this scalar field becomes a null state in four-dimensions. There are three important ingredients involved in this result: Gauge invariance, decoupling between transverse and longitudinal modes, and the functional form of $\mathcal{G}_\xi$.  From gauge invariance, both fields $\lambda$ and $\rho$ should be considered as physical, as they are gauge-independent. The decoupling condition forces their 4D profiles to be proportional. This means that their zero modes $\lambda_0$ and $\rho_0$ describe the same d.o.f. in four-dimensions, but now the localization of such a scalar zero mode requires the interplay of two different mode profiles instead of one. At this point, the tuning of the parameters in $\mathcal{G}$, imposed to ensure the existence of a zero-energy ground state in the gauge sector, seems to be the condition needed to remove the scalar zero modes from the low-energy spectrum. However, a better understanding on this phenomenon is still required. 
  
We can compare our results with two closely related  works: \cite{Zhao:2014iqa, Alencar:2014moa}.  Both papers treat the problem of gauge field localization by explicitly breaking the 5D gauge symmetry through the introduction of a geometrical coupling proportional to the scalar curvature $R$ that plays the role of a 5D mass. In our gauge-invariant analysis, their results for the gauge boson localization can be obtained straightforwardly upon the identification $\mathcal{G}_\xi(\phi)= \chi\mathcal{R}(\phi)$, with $\mathcal{R}$ defined in Eq.(\ref{Rphi}), which holds for the parameter values $\xi=1/2$ and $\chi=-1/16$.   

Finally, as stated in the introduction, our Stueckelberg-like mechanism is only suitable for abelian fields and a possible generalization to non-abelian fields is not straightforward. In fact the localization of non-abelian fields is an issue in many  (warped) compactifications as, naively, the cubic and quartic terms in the four-dimensional effective action get different couplings since the overlap integrals are different~\cite{Rubakov:2001kp}.  Within the RS1 scenario, a possible way-out was investigated in~\cite{Batell:2006dp} where a spontaneously broken 5D model was considered, and brane couplings were added to restore the four-dimensional gauge symmetry. However, it is also shown that additional scalar fields are needed to generate the necessary spontaneous symmetry breaking, and that the scalar mixing can in principle lead to strong coupling problems or quantum instabilities (ghosts).   Within the smooth brane scenario a possibility that is similar in spirit to the one considered here might be to take a scalar-field dependent gauge coupling~\cite{Kehagias:2000au}. Still, in both cases, charge universality of non-abelian gauge theories coupled to matter must be carefully addressed.


\section{Localization of the Kalb--Ramond field}\label{KRL}

The KR field can also be localized on the brane using an analogous procedure to the one presented in the previous section. We start with the 5D Stueckelberg-like formulation of the KR action 
\begin{eqnarray}
S_{KR}&=&\int d^5x \sqrt{-g}\left\{-\frac{1}{12}H^{MNL}H_{MNL}\right.\nonumber\\&&\left.-\frac{1}{4}\mathcal{F}(\phi)\left\{\partial_{\, \left[M\right. } C_{ \left. N\right]}-B_{MN}\right\}^2\right\}\,,  \label{KR1}
\end{eqnarray}
with the KR field strength defined as
\begin{equation}
H_{MNL}=\partial_M B_{NL}+\partial_L B_{MN}+\partial_N B_{LM}.  \label{KR2}
\end{equation}
Here $C_M$ plays the role of a Stueckelberg compensator and the function $\mathcal{F}(\phi)$ models the coupling between the KR field and the domain wall. Again, the back-reaction of $\mathcal{F}(\phi)$ to the geometry is neglected.

\subsection{Equations of motion and gauge fixing}
The action $S_{KR}$ in Eq.~(\ref{KR1}) is gauge-invariant under the transformation
\begin{equation}\label{gtKR1}
B_{MN} \rightarrow  B_{MN}+\partial_{\, \left[M\right. } \Lambda_{ \left. N\right]},\qquad
C_M \rightarrow C_M+ \Lambda_M,
\end{equation}
and its 5D equations of motion are
\begin{eqnarray}
&&\partial _{M}\Big(e^{4A}g^{MQ}g^{NR}g^{LS}H_{QRS}\Big)\nonumber\\&&\qquad=-e^{4A}\mathcal{F} g^{NR}g^{LS}\left\{\partial_{\, \left[R\right. } C_{ \left. S\right]}-B_{RS}\right\}\,,\label{5deomKR1}\\
&&\partial _{M}\Big\{e^{4A}\mathcal{F}g^{MQ}g^{NR}\left\{\partial_{\, \left[Q\right. } C_{ \left. R\right]}-B_{QR}\right\}\Big\}=0\,.\label{5deomKR2}
\end{eqnarray}

Parallel to the analysis of the previous section, we parameterize the 5D field $B_{MN}$ as
\begin{equation}
B_{MN}=\left( \begin{array}{c|c}
\widehat B_{\mu\nu}+\partial_\mu \varphi_\nu-\partial_\nu \varphi_\mu & {B}_{4\rho}\\
\hline
{B}_{\rho 4} & 0\\
\end{array}\right)\,,
\label{parKR}
\end{equation}
where $\widehat{B}_{\mu\nu}$ are the transverse components of ${B}_{\mu\nu}$ ($\partial^\mu \widehat{B}_{\mu\nu}=0$, $\widehat{H}_{\mu\nu\rho}=\partial_\mu \widehat{B}_{\nu\rho}+\partial_\nu \widehat{B}_{\rho\mu}+\partial_\rho \widehat{B}_{\mu\nu}$), while 
$\varphi_\mu$ stand for the corresponding vector components ($\partial_\mu \varphi_{\nu\rho}+\partial_\nu \varphi_{\rho\mu}+\partial_\rho\varphi_{\mu\nu}=0$, with $\varphi_{\mu\nu}=\partial_\mu \varphi_\nu-\partial_\nu \varphi_\mu$). Their behavior under the gauge transformation Eq.~(\ref{gtKR1}) is given by
\begin{equation}\label{gtKR2}
\begin{array}{l}
\widehat{B}_{\mu\nu} \rightarrow \widehat{B}_{\mu\nu}\,,\\
\varphi_\mu\rightarrow\varphi_\mu+\Lambda_\mu\,,\\
{B}_{4\mu} \rightarrow {B}_{4\mu}+ \partial_y\Lambda_{\mu}-\partial_\mu\Lambda_4.
\end{array}
\end{equation}
Again, upon integration by parts, the transverse field $\widehat{B}_{\mu\nu}$ decouples from the 
vector fields and the action $S_{KR}$ can be written as
\begin{eqnarray}
&&S_{KR}=S_{\widehat{B}}+S_V\,,\\
&&S_{\widehat{B}}=\int d^5 x\left\{ -\frac{1}{12}e^{-2A}\widehat{H}^2_{\mu\nu\rho}
-\frac{1}{4}(\widehat{B}'_{\mu\nu})^2-\frac{1}{4}\mathcal{F} 
\widehat{B}_{\mu\nu}^2\right\}\,,\nonumber\\\label{KRtens}\\
&&S_V=
   \int d^5x  \left\{ -\frac{1}{4}\lambda_{\mu\nu}^2
-\frac{1}{4}\mathcal{F}\rho_{\mu\nu}^2-\frac{1}{2}e^{2A}\mathcal{F} (\lambda_\mu-\rho'_\mu)^2 \right\}\,,\nonumber\\
\label{KRvec}
\end{eqnarray}
together with the field redefinitions
\begin{equation}\label{parKR2}
\begin{array}{rclcrcl}
\lambda_\mu&=& B_{4\mu}-\varphi'_\mu+\partial_\mu C_4,&\qquad&\rho_\mu&=&C_\mu-\varphi_\mu\,,\\
\lambda_{\mu\nu}&=&\partial_\mu\lambda_\nu-\partial_\nu\lambda_\mu,&\qquad&\rho_{\mu\nu}&=&\partial_\mu\rho_\nu-\partial_\nu\rho_\mu,
\end{array}
\end{equation}
where the new fields $\lambda_\mu$ and $\rho_\mu$ are now invariant under Eqs.~(\ref{gtKR1}), (\ref{gtKR2}):
\begin{equation}\label{gtKR3}
\lambda_\mu\rightarrow \lambda_\mu,\qquad
\rho_\mu\rightarrow \rho_\mu.
\end{equation}

In terms of the gauge-invariant fields $\widehat{B}_{\mu\nu}$, $\lambda_\mu$, and $\rho_\mu$, Eqs.~(\ref{5deomKR1}), (\ref{5deomKR2}) read
\begin{eqnarray}
&&\left[e^{-2A}\square+\left(\partial^2_y-\mathcal{F}\right)\right]\widehat{B}_{\mu\nu}=0\, ,  \label{eomKR1}\\
&& \lambda'_{\mu}-\mathcal{F}\rho_{\mu}=0\, , \label{eomKR4}\\
&&\partial ^{\mu}\lambda_{\mu\nu}+e^{2A}\mathcal{F}\left(\rho'_\nu-\lambda_\nu\right)=0\, ,  \label{eomKR2}\\
&&\mathcal{F}\partial ^{\mu}\rho_{\mu\nu}+\partial_y\left[e^{2A}\mathcal{F}_\kappa\left(\rho'_\nu-\lambda_\nu\right)\right]=0\, .  \label{eomKR3}
\end{eqnarray}
Notice here that Eqs.~(\ref{eomKR1},\ref{eomKR4}) follow from
\begin{equation}
\left[e^{-2A}\square+\left(\partial^2_y-\mathcal{F}\right)\right]\widehat{B}_{\mu\nu}=\lambda'_{\mu\nu}-\mathcal{F}\rho_{\mu\nu}\,, \label{eomKR0}
\end{equation}
after isolating its transverse and vector parts, while  Eq.~(\ref{eomKR4}) is  again satisfied in a weaker form by a combination of the Eqs.~(\ref{eomKR2},\ref{eomKR3}):
\begin{equation}
\partial^\mu\left[\lambda'_{\mu\nu}-\mathcal{F}\rho_{\mu\nu}\right]=0\,.\label{eomKR5}
\end{equation}

\subsection{Brane coupling}

Decomposing the antisymmetric field  $\widehat{B}^{\mu\nu }$ as
\begin{equation}
\widehat{B}^{\mu\nu }=\sum_{n}b_{n}^{\mu\nu }(x)\eta_{n}(y), \label{decKR1}
\end{equation}
Equation (\ref{eomKR1}) becomes
\begin{equation}
\eta''_n(y)-\mathcal{F}\eta_n(y)=-e^{-2A}M^2_n\eta_n(y).  \label{profKR1}
\end{equation}
Taking an analogous course of action as that for the gauge vector case, we propose as an Ansatz for $\mathcal{F}(\phi)$ the following uni-parametric family of functions that admit a normalizable zero-energy ground state:
\begin{align}
\mathcal{F}(\phi)&=\mathcal{F}_\kappa(\phi)\nonumber\\&\equiv-\left(\frac{\kappa+1}{6}\right)\left\{[\partial_\phi W(\phi)]^2-\left(\frac{\kappa+1}{6}\right) W(\phi)^2\right\}\,,  \label{F1}
\end{align}
or equivalently
\begin{equation}
\mathcal{F}_\kappa[\phi(y)]=(\kappa+1) \left[(\kappa+1)[A'(y)]^2+ A''(y)\right],  \label{F2}
\end{equation}
with real positive parameter $\kappa$. Inserting this functional into Eq.~(\ref{profKR1}) the mode profiles are then determined by
\begin{align}
&\eta''_n(y)-(\kappa+1) \left[(\kappa+1)[A'(y)]^2+ A''(y)\right]\eta_n(y)\nonumber\\&\qquad=-e^{-2A}M^2_n\eta_n(y),  \label{profKR2}
\end{align}
or in conformally flat space-time coordinates
\begin{equation}
\left[-\partial_z^2+U_{\kappa}(z)\right]\theta_n(z)=Q_{\kappa}^\dagger Q_{\kappa}\theta_n(z)=M^2_n\theta_n(z),  \label{profKR3}
\end{equation}
with
\begin{equation}
\eta_n(z)=e^{A(z)/2}\theta_n(z).
\end{equation}

Now the zero-energy wavefunction annihilated by $Q_{\kappa}$ is  
\begin{equation}
\theta_0(z)\propto e^{(\kappa+\frac12)A(z)},  \label{wfzmKR1}
\end{equation}
and the corresponding zero mode profile $\eta_0(y)$ becomes
\begin{equation}
\eta_0(y)\propto e^{(\kappa+1) A(y)},\label{pzmsolKR1}
\end{equation}
which is normalizable for $\kappa>0$\,.

\subsection{Vector Sector}

The fate of the vector sector can be studied decomposing the vector fields as
\begin{equation}
\lambda^\mu=\sum_{n}\lambda_{n}^\mu(x)u_{n}(y), \qquad
\rho^\mu=\sum_{n}\rho_{n}^\mu(x)v_{n}(y),\label{decKR2}
\end{equation}
and defining the modes as
\begin{equation}
\partial_\mu\lambda^{\mu\nu}_n	=M_{Vn}^2\lambda_{n}^\nu,\qquad
\partial_\mu\rho^{\mu\nu}_n	=M_{Vn}^2\rho_{n}^\nu.  \label{modKR2}
\end{equation}
From Eq.~(\ref{eomKR2}), the vector zero modes satisfy
\begin{equation}
\lambda_{0}^\mu(x)u_{0}(y)=\rho_{0}^\mu(x)v'_{0}(y). \label{zmvecKR}
\end{equation}
Substituting this relation into Eq.~(\ref{eomKR4}), one has 
\begin{equation}
v''_{0}(y)-\mathcal{F}_\kappa v_{0}(y)=0.
\end{equation}
Thus, for $\kappa>0$, there is only one normalizable solution, given by 
\begin{equation}
v_0(y)\propto e^{(\kappa+1) A},\label{pzmsolKR2}
\end{equation}
but again this mode fails to be localized in the brane, as its effective action vanishes upon integration over the extra dimension:
\begin{equation}
S^{0}_V= \int d^4x \left[-\frac{1}{4}\left(\rho^{\mu\nu}_{0}\right)^2 \right] \int_{-\infty}^{+\infty} dy\, \left[ (v'_{0})^2+\mathcal{F}_\kappa v_{0}^2\right]=0\,.
\label{LeKR}
\end{equation}
Thus, in this case there is also a null state in the low-energy spectrum, whose properties must be studied in detail.

On the other hand,  from Eq.~(\ref{eomKR4}) the following relation holds for all modes:
\begin{equation}\label{subcKR3}
\mathcal{F}_\kappa\rho^\mu_{n}v_n=\lambda^\mu_{n}u'_n.
\end{equation}
Plugging this relation into Eq.~(\ref{eomKR2}) we obtain the definitive equation for the mode profiles
\begin{equation}
\mathcal{F}_\kappa u''_n(y)
-\mathcal{F}'_\kappa u'_n(y)+\mathcal{F}_\kappa\left(e^{-2A}M_{Vn}^2-\mathcal{F}_\kappa\right)u_{n}(y)
=0.  \label{profzmKR3}
\end{equation}

As a summary of the results obtained in this section, we can state that the model defined by Eqs.~(\ref{KR1}), (\ref{F1}) contains a normalizable zero-energy ground state described by a massless 2-form in four dimensions and a continuous tower of massive 2-forms and 1-forms.  The spectrum also includes a massless 1-form, which is again a null state in the 4D effective low-energy action.

Before closing this section, let us point out a possible natural connection between the functions $\mathcal{G}_\xi$ and $\mathcal{F}_\kappa$. If we require the brane--gauge coupling to be universal, imposing
\begin{equation}\label{univ1}
\frac{\mathcal{G}_\xi}{2}=\frac{\mathcal{F}_\kappa}{4}
\end{equation}
in Eqs.~(\ref{2},\ref{KR1}), then, from Eqs.~(\ref{G4}), (\ref{F1}) the only solution of Eq.~(\ref{univ1}) satisfying the conditions $\xi>0$ and $\kappa>0$ is $\xi=2$, $\kappa=3$. This particular choice renders functions which are simply proportional to the scalar potential $V(\phi)$:
\begin{equation}
\frac{1}{2}\mathcal{G}_2[\phi(y)]=\frac{1}{4}\mathcal{F}_3[\phi(y)]=-\frac{1}{3}V(\phi). 
\end{equation}
Thus, in this special case, the localization of abelian gauge fields is driven by the very same function that determines the background geometry.

\section{Conclusions}\label{Conc}

We have proposed a new mechanism for abelian gauge field localization on thick branes. The key feature of the model is the presence of Stueckelberg-compensating fields, which allow for the introduction of Yukawa-like interactions in a gauge-invariant (and Einstein-covariant) way. 

In the vector case, the interaction between the brane and the gauge field is modeled by a function $\mathcal{G}(\phi)$ that depends on the classical background responsible for the brane formation. Identifying the brane with a domain-wall solution generated by a single real scalar $\phi$ through some super-potential $W(\phi)$, with asymptotic AdS behavior in five dimensions, we have shown that there is a whole family of functions $\mathcal{G}_\xi(\phi)$ ---constructed from $W(\phi)$ and its derivatives--- which guarantee the existence of a normalizable zero-energy vector ground state in the theory. We have also studied the scalar sector of the model, concluding that despite being apparently normalizable, the scalar zero mode is not trapped by the brane. It turns out to be a null state in the effective 4D low-energy theory. The presence of such a null state  might indicate either the presence of a hidden left-over gauge symmetry or the presence of a quantum instability (a ghost mode) in the bulk spectrum.  In fact it is interesting to notice how the cosmological setup of~\cite{Herrera-Aguilar:2014oua} also displays an instability, of tachyonic type. It would be interesting to investigate further on these aspects.  

The same localization mechanism can be straightforwardly applied to the abelian KR field. In this case, we have also found a one-parameter family of brane--gauge coupling functions $\mathcal{F}_\kappa$ compatible with the presence of a normalizable zero mode. Such a straightforward generalization is expected to be kept for a generic antisymmetric form field.   

Finally, we have shown that if the brane--gauge coupling is universal, then it must be proportional to the scalar potential $V(\phi)$, the same function that triggers the brane formation.

\begin{acknowledgements}
This work was partly supported by CONACyT under project CB-2011-167425 and by UCMEXUS-CONACyT grant CN-12-564. The authors would like to thank A.~Aranda for help and suggestions and for participating in the earlier stages of this project, and Z.~Kakushadze for careful reading an earlier version of the manuscript. CAV-A acknowledges the Mainz Institute for Theoretical Physics (MITP) for hospitality and support during the revision stage of this work.
\end{acknowledgements}

\end{document}